\documentclass[rnote, longauth]{aa} 

\usepackage{ulem}
\usepackage{graphicx}
\usepackage{comment}
\usepackage{txfonts}
\usepackage{ctable}
\usepackage{hyperref}
\usepackage{xspace}
\usepackage{epstopdf, epsfig}
\usepackage{tabularx}
\usepackage{ulem}

\newcommand\g{\ensuremath{\gamma}}%
\newcommand\hess{H.E.S.S.\xspace}
\newcommand\fermi{\textit{Fermi}-LAT\xspace}

\newcommand\snrG{\object{G349.7$+$0.2}\xspace}
\newcommand\PA{\object{Puppis A}\xspace}

\def\gr{$\gamma$-ray}
\def\grs{$\gamma$-rays}

\def\eg{{e.g.~}}

\begin{document} 

   \title{H.E.S.S. reveals a lack of TeV emission from the supernova remnant Puppis A}

\author{\small H.E.S.S. Collaboration
\and A.~Abramowski \inst{1}
\and F.~Aharonian \inst{2,3,4}
\and F.~Ait Benkhali \inst{2}
\and A.G.~Akhperjanian \inst{5,4}
\and E.O.~Ang\"uner \inst{6}
\and M.~Backes \inst{7}
\and S.~Balenderan \inst{8}
\and A.~Balzer \inst{9}
\and A.~Barnacka \inst{10,11}
% ...
}

\institute{
Universit\"at Hamburg, Institut f\"ur Experimentalphysik, Luruper Chaussee 149, D 22761 Hamburg, Germany \and
Max-Planck-Institut f\"ur Kernphysik, P.O. Box 103980, D 69029 Heidelberg, Germany \and
Dublin Institute for Advanced Studies, 31 Fitzwilliam Place, Dublin 2, Ireland \and
National Academy of Sciences of the Republic of Armenia,  Marshall Baghramian Avenue, 24, 0019 Yerevan, Republic of Armenia  \and
Yerevan Physics Institute, 2 Alikhanian Brothers St., 375036 Yerevan, Armenia \and
Institut f\"ur Physik, Humboldt-Universit\"at zu Berlin, Newtonstr. 15, D 12489 Berlin, Germany \and
University of Namibia, Department of Physics, Private Bag 13301, Windhoek, Namibia \and
University of Durham, Department of Physics, South Road, Durham DH1 3LE, U.K. \and
GRAPPA, Anton Pannekoek Institute for Astronomy, University of Amsterdam,  Science Park 904, 1098 XH Amsterdam, The Netherlands \and
Obserwatorium Astronomiczne, Uniwersytet Jagiello{\'n}ski, ul. Orla 171, 30-244 Krak{\'o}w, Poland \and
now at Harvard-Smithsonian Center for Astrophysics,  60 Garden St, MS-20, Cambridge, MA 02138, USA
% ...
}

   \authorrunning{The H.E.S.S.~collaboration}
   \titlerunning{Lack of TeV emission from Puppis A}
   
    \offprints{Diane Fernandez\\
      \email{diane.fernandez@lupm.univ-montp2.fr}\\
      Igor Oya\\
      \email{igor.oya.vallejo@desy.de}
    }

   \date{Received 14 August 2014; accepted DAY MONTH 2014}

   \abstract{ {\it Context:} \PA is an interesting $\sim$4 kyr-old
     supernova remnant (SNR) that shows strong evidence of interaction
     between the forward shock and a molecular cloud. It has been
     studied in detail from radio frequencies to high-energy (HE,
     0.1-100 GeV) \g-rays. An analysis of the \fermi data has shown
     extended HE \g-ray emission with a 0.2-100 GeV spectrum
     exhibiting no significant deviation from a power law, unlike most
     of the GeV-emitting SNRs known to be interacting with molecular
     clouds. This makes it a promising target for imaging atmospheric
     Cherenkov telescopes (IACTs) to probe the \g-ray emission above
     100 GeV. \\ {\it Aims:} Very-high-energy (VHE, E $\ge$ 0.1 TeV)
     \g-ray emission from \PA has been, for the first time, searched for with
     the High Energy Stereoscopic System (H.E.S.S.).\\ {\it Methods:}
     Stereoscopic imaging of Cherenkov radiation from extensive air
     showers is used to reconstruct the direction and energy of the
     incident \g-rays in order to produce sky images and source
     spectra. The profile likelihood method is applied to find constraints on
     the existence of a potential break or cutoff in the photon
     spectrum.\\ {\it Results:} The analysis of the H.E.S.S. data does
     not reveal any significant emission towards \PA. The derived
     upper limits on the differential photon flux imply that its
     broadband \g-ray spectrum must exhibit a spectral break or
     cutoff. By combining \fermi and H.E.S.S. measurements, the 99\%
     confidence-level upper limits on such a cutoff are found to be
     450 and 280 GeV, assuming a power law with a simple exponential
     and a sub-exponential cutoff, respectively. It is concluded that
     none of the standard limitations (age, size, radiative losses) on
     the particle acceleration mechanism, assumed to be continuing at present,
     can explain the lack of VHE signal. The scenario in which particle acceleration has ceased some time ago is considered as an alternative explanation. The HE/VHE spectrum of \PA
     could then exhibit a break of non-radiative origin (as observed
     in several other interacting SNRs, albeit at somewhat higher
     energies), owing to the interaction with dense and neutral
     material, in particular towards the NE region.}


   \maketitle
%

\section{Introduction}
Supernova remnants (SNRs) have long been considered as the main
sources of Galactic cosmic rays
\citep[CRs,][]{1964SvA.....8..342G}. Direct measurements of CRs from
SNRs are impossible because of Galactic magnetic fields, but \grs\, can
provide an indirect signature of their presence \citep[see \eg][for a
review]{2008ARA&A..46...89R}. On one hand, several middle-aged SNRs
interacting with molecular clouds (MCs) have been observed with the
{\it Fermi} Large Area Telescope (\fermi) and Astro-Rivelatore Gamma a
Immagini Leggero ({\it AGILE}) telescope as luminous high-energy (HE,
0.1$-$100 GeV) \g-ray sources. The strong HE \g-ray emission from
these sources is thought to arise from neutral pion decay subsequent
to the interactions between accelerated CR particles and the dense
gas, and whose unique spectral feature, referred to as the pion bump
at $\sim$ 200 MeV, was recently revealed in two such SNRs with \fermi
\citep{2013Sci...339..807A}.  Some examples are \object{W28}
\citep{2010ApJ...718..348A}, \object{W51C}
\citep{2009ApJ...706L...1A}, W44 \citep{2010Sci...327.1103A}, and
\object{IC 443} \citep{2010ApJ...712..459A}, which furthermore all
exhibit power-law spectral breaks in the 1$-$20 GeV range. As a
consequence of these breaks, the associated very-high-energy (VHE, E
$\ge$ 0.1 TeV) emission is usually soft and faint
\citep[\eg][]{2007ApJ...664L..87A,2008A&A...481..401A}. On the other
hand, young, shell-type SNRs, with bright and hard spectra in the VHE
domain \citep[such as \object{RX
  J1713.7$-$3946},][]{2007A&A...464..235A} and without clear evidence
for cloud interaction, exhibit hard and relatively faint spectra in
the HE domain \citep{fermi_2011}. In these cases, inverse Compton (IC)
emission from accelerated electrons naturally explains the observed
\gr\ emission \citep[\eg][]{lee_2012}.

\PA (G260.4$-$3.4) represents an interesting case in between these two SNR categories. At a distance of 2.2 $\pm$ 0.3
kpc\footnote{Although a smaller distance of 1.3 kpc, with large
  uncertainties of +0.6/-0.8 kpc, has been previously determined by
  \citet{2000MNRAS.317..421W} based on OH line observations, a
  distance of 2.2 kpc is assumed throughout this paper.}
\citep{2003MNRAS.345..671R}, it is a well-studied Galactic SNR in most
energy bands from radio to HE \grs. It is one of the three
oxygen-rich SNRs \citep{1985ApJ...299..981W} known today in the
Galaxy. This, together with the presence of a central compact object
\citep[CCO;][and references therein]{2012ApJ...755..141B}, strongly
supports the idea that \PA originates in a core-collapse SN
explosion. Based on the motions of both optical filaments and CCO, its
age is estimated to be (4450$~\pm$~750)~yr
\citep{2012ApJ...755..141B}, implying that the SNR is currently in the
Sedov-Taylor evolutionary phase (see \eg
\citealt{1977ARA&A..15..175C}).  The strong X-ray emission from \PA is
mostly dominated by the shock-heated interstellar medium \citep[ISM,
\eg][]{2005ApJ...635..355H}, except for some isolated O-Ne-Mg-rich
features associated with the SN ejecta
\citep{2008ApJ...676..378H,2008ApJ...678..297K,2010ApJ...714.1725K},
the kinematics of which have been fully measured \citep[][and
references therein]{2013ApJ...768..182K}. The SNR also has another
interesting characteristic: together with \object{W49B}
\citep{2010ApJ...722.1303A} and \snrG \citep{2010MNRAS.409..371L}, it
is amongst the youngest Galactic SNRs known to be interacting at
several locations throughout the shell with dense gas seen as a
complex of $\ion{H}{I}$ and CO clouds surrounding most of the SNR
\citep{1988A&AS...75..363D,1995AJ....110..318R,2013A&A...555A...9D}. In
particular, spectro-imaging X-ray studies towards the so-called bright
eastern knot (BEK) have presented evidence for a shock-cloud
interaction \citep{2005ApJ...635..355H}. Recent high-resolution X-ray
observations of the whole SNR \citep{2013A&A...555A...9D} have
confirmed the presence of a decreasing gradient in the emission from
NE to SW and also revealed a highly structured and
filamentary morphology with unprecedented detail, indicating that \PA
is evolving in an inhomogeneous, knotty ISM. Observations with {\it
  Spitzer} have shown a clear correlation between infrared (IR) and
X-rays at all spatial scales, demonstrating that the thermal IR
emission arises from dust collisionally heated by the hot, shocked
plasma \citep{2010ApJ...725..585A}.

\cite{2012ApJ...759...89H} have reported the detection of \PA in the
HE \g-ray domain with the \fermi. Its luminosity of 2.7 $\times$
10$^{34}$(d/2.2 kpc)$^2$ erg s$^{-1}$ in the 1$-$100 GeV band is
slightly higher than those of the low-luminosity, HE-emitting SNRs
such as \object{Cygnus Loop}, \object{S147} and \object{HB21}
(\citealt{2012A&A...546A..21R} and references therein), and about a
factor ten lower than those measured from the archetypal SNRs known to
be interacting with MCs
\citep{2009ApJ...706L...1A,2010ApJ...712..459A,2010ApJ...722.1303A}. The
morphology of the HE \gr\, emission is described well by a uniform
disc of radius 0.38$^{\circ}\pm0.04^{\circ}$ and compatible with the
radio and X-ray morphologies. The HE \gr\, spectrum is described
well by a power law (PL) with no indication of a break or cutoff, and a
spectral variation at the $\sim$2$\sigma$ level between the E
and the W hemispheres was found
.
Such a PL HE spectrum, together with a hint of a radio break at
$\sim$40 GHz in the WMAP data \citep{2012ApJ...759...89H}, makes \PA
quite peculiar with respect to most of the HE-emitting SNRs. \PA can
be considered as an intermediate case between the young, isolated, and
bright VHE \g-ray emitting SNRs and the middle-aged, bright HE \g-ray
emitting ones interacting with MCs.
To probe its emission in the VHE domain, observations towards \PA
obtained with the High Energy Stereoscopic System (H.E.S.S.) are
reported in this article.

\section{Observations and analysis} 
\subsection{\hess observations and analysis results}
\label{sec-hess}

\hess is an array of five imaging atmospheric Cherenkov telescopes
(IACTs) located in Namibia and designed to detect VHE
\grs~\citep{2003APh....20..111B}. The fifth telescope (28-m diameter)
has been operating since September 2012, but the data exploited here
were taken from 2005 to 2013 with the four-telescope array
alone. In this configuration, the instrument covers a field of view of
5\degr. The primary particle direction and energy are reconstructed
above a threshold of $\sim$100 GeV with an angular resolution of
$\sim$0.1\degr~and energy resolution of $\sim$15\%,
\citep[\eg][]{2006A&A...457..899A}. The whole dataset on \PA amounts
to 24\,h including 17\,h of dedicated observations taken using the
{\it wobble} mode and 7\,h of runs towards nearby sources in the Vela
region \citep[in particular,
\object{Vela~X},][]{2006A&A...448L..43A,2012A&A...548A..38A}. These
observations were performed at zenith angles between $18^{\circ}$ and
$45^{\circ}$ with a median value of $22^{\circ}$, and a median offset
from the source of $1.0^{\circ}$.

\begin{figure}[!htb]
\centering
\includegraphics[width=\columnwidth]{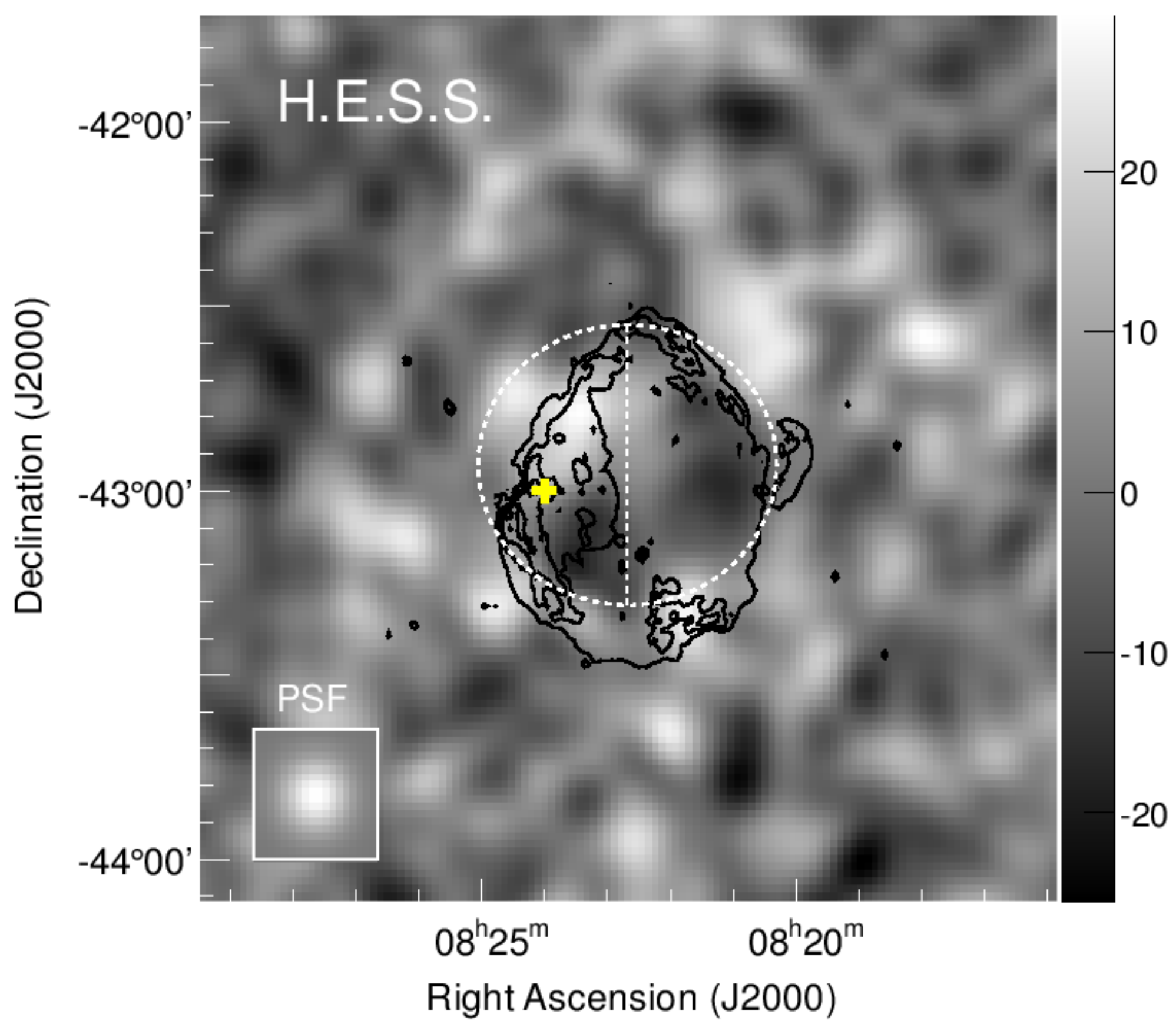}

\caption{Image of the \hess~\gr\, excess centred on the \PA SNR. The
  excess was smoothed with a Gaussian kernel of width $0.06\degr$
  corresponding to the \hess angular resolution (68\% containment
  radius) for this analysis (shown in the bottom left inset). The
  colour scale represents the excess counts per surface area of $\pi
  (0.06\degr)^{2}$. The circular analysis region of radius
  $0.38\degr$, matching the \fermi best-fit morphological model, is
  shown as a dashed circle. The two hemispheres are separated by a
  dashed line along the north/south axis in celestial coordinates. The
  black contours represent the 1.4 GHz continuum emission
  \citep{2006A&A...459..535C} at the 5, 10, 20 and 50 mJy/beam
  levels. The yellow cross indicates the position of the BEK
  \citep{2005ApJ...635..355H}.}

\label{fig:map}
\end{figure}

Data were analysed with the model analysis described in
\citet{2009APh....32..231D} and using \textit{Standard
  cuts}\footnote{The H.E.S.S. {\it ParisAnalysis version 0-8-24}
  software with {\it Prod26 DSTs} was used.}. The main analysis
results were confirmed with an independent data calibration chain and
a multivariate analysis method \citep{2009APh....31..383O}. The
resulting energy threshold of the main analysis, which is conservatively
defined as the energy above which the acceptance is larger than 15\%
of its maximum value, is $E_\mathrm{th}=0.26$\,TeV. The analysis
ON-region (i.e. signal integration region) was defined as a circular
region of radius $0.38^{\circ}$, centred on
$\alpha_\mathrm{J2000}=08^\mathrm{h} 22^\mathrm{m} 40\fs8$,
$\delta_\mathrm{J2000}=-42\degr 55\arcmin 48\farcs0$, to match the
best-fit values of the HE \g-ray emission extent and position measured
with the \fermi\footnote{The HE spectral parameters were obtained
  with the ROSAT X-ray template.  However, as shown in Table 3 of
  \cite{2012ApJ...759...89H}, those derived under the assumption of a
  uniform disc are fully compatible.}.  The statistical significance
of a potential VHE \g-ray emission from \PA\ was determined by using
equation (17) in \citet{1983ApJ...272..317L} after background
subtraction with the reflected background method
\citep{2007A&A...466.1219B}.  No significant signal was found within
the ON-region. In total, eight excess counts were measured, corresponding
to a significance of $0.1 \sigma$. Similar analyses have been
performed for two half-disc regions corresponding to the E and
W hemispheres as defined in \citet{2012ApJ...759...89H}. Figure
\ref{fig:map} shows an image of the \g-ray excess counts, where the
background level is estimated following the template background method
and cross-checked with the ring background method
\citep{2007A&A...466.1219B}, together with the ON-region and contours
of the radio continuum emission at 1.4~GHz
\citep{2006A&A...459..535C}.

\begin{figure} [h]
\centering
\includegraphics[width=\columnwidth]{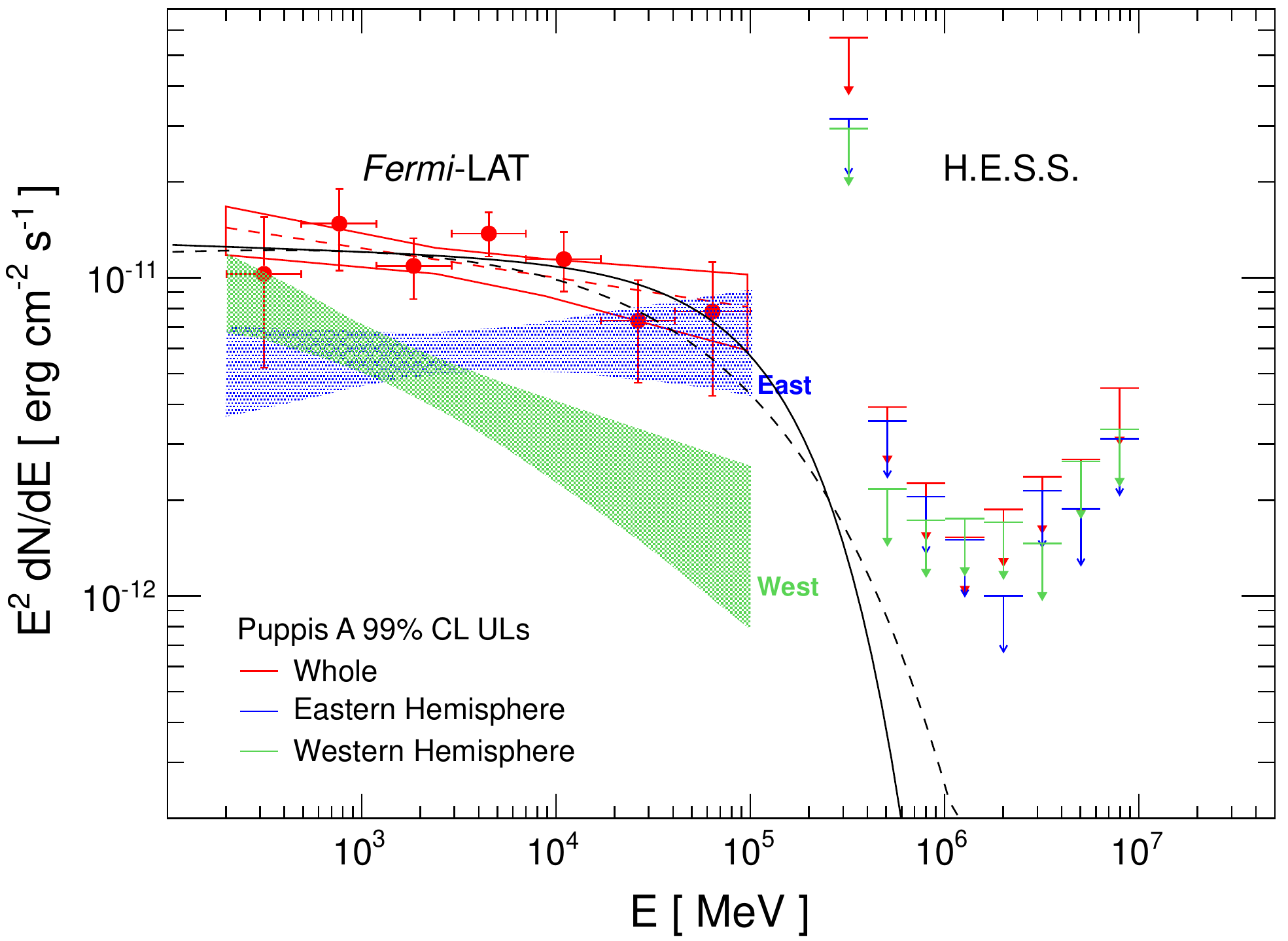}

\caption{\hess 99\% CL upper limits on the differential flux (arrows),
  together with the \fermi spectra from \PA, as reported in
  \citet{2012ApJ...759...89H}. Red, blue and green symbols correspond
  to \fermi~and \hess~measurements for the whole SNR, the E and
  W hemisphere, respectively. The data points show the LAT
  fluxes and 1 $\sigma$ statistical and systematic errors, whilst the
  bow-tie areas define the 68\% CL bands. The solid and dashed lines
  indicate the preferred \g-ray spectra for the exponential and
  sub-exponential cutoff models, respectively.}
\label{fig:UL}
\end{figure}

Following the method of \citet{1998PhRvD..57.3873F}, differential flux
upper limits (ULs) at the 99\% confidence level (CL) and for a
spectral index $\Gamma\,$=\,2.1 \citep[as measured with
\fermi,][]{2012ApJ...759...89H} were extracted in the 0.26$-$10 TeV
energy range within the circular and the two half-disc
ON-regions. These ULs are not very sensitive to the choice
of the photon index; assuming $\Gamma$ = 3 instead changes the values
by less than 5\%. To ensure that these ULs account for the full source
emission, they have been corrected for the underestimation caused by
events reconstructed outside the analysis regions due to the
\hess~point spread function (PSF). By convolving a disc of
$0.38^{\circ}$ radius with the \hess~PSF, the flux outside of the
ON-region was estimated to be 10\% of the total flux. The H.E.S.S. ULs
do not vary by more than 10\% by changing the integration radius
between 0.35$^{\circ}$ and 0.48$^{\circ}$. The resulting \hess~ULs are
shown in Fig. \ref{fig:UL} together with the \fermi spectra.

While the \hess~ULs derived from the W hemisphere are not
constraining, owing to a steeper HE spectrum, those from the E
hemisphere and from the whole SNR exclude the possibility that their
respective PL spectra extend up to the VHE domain at more than the
99\% CL. This indicates the existence of a spectral feature (break or
cutoff) at intermediate energies, i.e. between the HE~and VHE~domains.

\subsection{Constraints on the CR particle spectrum}
\label{sec:RatioTest}

Presuming that the accelerated particle spectra are PLs with
exponential cutoffs as predicted by the diffusive shock acceleration
(DSA) mechanism (e.g.  \citealt{1977SPhD...22..327K,
  1978ApJ...221L..29B, 1978MNRAS.182..147B, 1978MNRAS.182..443B}), the
\g-ray spectrum resulting from the different emission processes should
also follow a PL with exponential cutoff: $dN/dE = N_0 E^{-\Gamma}
e^{-(E/E_{\rm cut})^{\beta}}$, where $E_{\rm cut}$ is the cutoff
energy and $\beta$ defines the spectral shape in the cutoff region. To
evaluate the existence of such a cutoff energy in the spectrum of the
whole SNR\footnote{Since no \fermi spectral points are available for
  the hemispheres, no ULs on a cutoff or break have been derived.}, a
likelihood estimator $\mathcal L$ was defined as the combination of
the likelihoods from the \fermi data points and
\hess~measurements. The \fermi likelihood value was estimated by
computing $\chi^2$ from the available data points and 1$\sigma$
errors. The \hess~likelihood is calculated by comparing the number of
detected excess events (following Poisson statistics) with the
expected number in each reconstructed energy bin. To cover a wide
range of physically possible scenarios, two values for the $\beta$
parameter were chosen: $\beta$ = 1 (defining the \textit{exponential cutoff model} hereafter) corresponds to the most commonly
used case  and $\beta$ = 0.5 (defining the \textit{sub-exponential cutoff} model hereafter) is more physically motivated in both
leptonic (under the assumption that electrons suffer from SC losses
and that diffusion proceeds in Bohm regime,
\citealt{2007A&A...465..695Z}) and hadronic (for a proton spectrum
exhibiting a simple exponential cutoff, \citealt{2006PhRvD..74c4018K})
scenarios. The method used to derive the exclusion domain of $E_{\rm
  cut}$ is based on a likelihood ratio test statistic
\citep{2005NIMPA.551..493R}:
\begin{equation}
    \label{eq:lnL}
           \Lambda (E_{cut_0})  = \frac{\underset{\theta} \sup_{}\, \mathcal L(E_{cut_0}, \theta)} {\underset{E_{cut},\theta} \sup_{}\, \mathcal L(E_{cut}, \theta)},
\end{equation}
where $E_{cut_0}$ is the tested hypothesis and $E_{cut}$ all the
allowed values. The unknown spectral index $\Gamma$ and normalisation
$N_0$ are considered as nuisance parameters under the $\theta$
variable. The profile of the log-likelihood ratio test statistic
$-$2\,ln\,$\Lambda$ has an approximate $\chi^2$ distribution with one
degree of freedom \citep{2005NIMPA.551..493R}. The minimum is reached
at around $\sim$55~GeV and 150~GeV for $\beta$ = 0.5 and 1,
respectively. Below these energies, $-$2\,ln\,$\Lambda$ increases
rapidly because of the constraints imposed by the \fermi
detection. Above that, the \hess~data become more constraining and
lead to an increase in the $-$2\,ln\,$\Lambda$ value.  The 99\% CL ULs
on the cutoff energy correspond to 280 and 450~GeV for $\beta$ = 0.5
and 1, respectively\footnote{A toy model was used to study the
  coverage of this method by using Monte Carlo simulations. The
  results of these simulations indicate that the coverage is indeed
  fulfilled.}.

Broken PL spectra have been observed in several SNRs interacting with
MCs (\eg \citealt{2010ApJ...718..348A,2010Sci...327.1103A}), and this
could also be the case for \PA. A spectral index variation
$\Delta\Gamma$ $\sim$ 1 can be explained by radiative cooling of
electrons or escape of protons in the case of a SNR encountering a
dense and neutral medium \citep{2012PhPl...19h2901M}.  The derived UL
on the sub-exponential cutoff energy (280 GeV) can be used as a
conservative UL on the energy of any spectral break as long as
$\Delta\Gamma\lesssim$1.

\section{Discussion}
\label{sec-discu}

Throughout the SNR evolution, particles are accelerated at the forward
shock up to a maximum energy $E_{\rm max}$, typically determined by
the SNR's finite age, finite size, or radiative losses.  These effects
become relevant when the characteristic timescales are close to
the acceleration timescale, leading to cutoffs in the spectra of
accelerated particles residing in the SNR.  Additionally, a radiative
spectral break can be present at the particle energy $E_{\rm
  break}$ for which the radiative loss timescales equal the SNR age.
The conservative ULs on the cutoff energy in the $\gamma$-ray spectrum
derived from the \fermi and H.E.S.S. measurements translate into ULs
on E$_{\rm max}$ of $\sim$2\,TeV, $\sim$3\,TeV, or $\sim$5\,TeV
depending on whether the $\gamma$-ray emission results from
Bremsstrahlung (Br), IC or proton-nucleus interactions (e.g. p-p)
radiation mechanisms.  These limits can in turn be compared with the
expectations from DSA theory taken in its simplest form and applied to
the case of \PA whose main parameters are an age of $\sim$4500 yr
\citep{2012ApJ...755..141B}, a shock radius $R_{\rm sh}$=15~pc (for
$R_{\rm sh}$=24\arcmin~at 2.2 kpc), and a shock velocity v$_{\rm sh}$
ranging from 700 to 2500~km$\,\rm s^{-1}$. The last velocity value
was estimated by \cite{2013ApJ...768..182K} based on the electron
temperatures and ionisation timescales in the ejecta knots, whilst the
former was derived from the shock temperature of $\sim$0.7~keV
\citep{2008ApJ...676..378H} by assuming full equipartition of the
shock energy between ions and electrons. These shock-velocity
estimates concern the NE region of the SNR, which is
also the region coincident with the bulk of GeV emission. Based on
these SNR parameters, constraints on the magnetic field and ISM
density can be derived and compared to independent estimates
\citep{2012ApJ...759...89H,2013A&A...555A...9D}.  In these
calculations, the acceleration is assumed to proceed in a steady-state
manner from the SN until now, and the derived contraints are
implicitely considered as having been constant. Assuming Bohm
diffusion, the acceleration timescale\footnote{By setting the coefficient $k_{\rm 0}$ from Eq. (14) of \citep{2006A&A...453..387P} equal to unity.}  and the diffusion coefficient
are $\tau_{\rm acc} = (3\times10^3\, \mathrm{yr})\, E_{\rm TeV}\,
B_{\rm \mu G}^{-1}\,$ v$_{\rm sh,3}^{-2}\,$and $D(E_{\rm TeV}) =
(3.3\times10^{25}\, \mathrm{cm^{2}\, s^{-1}})\, E_{\rm TeV}\, B_{\rm
  \mu G}^{-1}\,$ \citep{2006A&A...453..387P}, where $B_{\rm \mu G}$ is
the downstream magnetic field in units of $\rm \mu G$, $E_{\rm TeV}$
the particle energy in units of TeV, and v$_{\rm sh,3}$ the shock
velocity in units of 10$^3\,\rm km\,s^{-1}$.  Synchrotron (syn), p-p,
and Br radiative loss timescales are $\tau_{\rm syn}\simeq
(2.1\times10^7\, \mathrm{yr})\, B_{\rm \mu G}^{-2}\, E_{\rm TeV}^{-1}$
\citep{2006A&A...453..387P}, $\tau_{\rm p-p}\simeq (5.3\times10^7\,
\mathrm{yr})\, n_{\rm 0}^{-1}$, and $\tau_{\rm br}\simeq
(3.3\times10^7\, \mathrm{yr})\, n_{\rm 0}^{-1}$
\citep{2009MNRAS.396.1629G}, with $n_{\rm 0}$ the ISM density in units
of cm$^ {-3}$, and $B_{\rm \mu G}$ is
the downstream magnetic field in units of $\rm \mu G$.
The derived constraints are shown in Table \ref{tab:constraints}.

\begin{table} [!htb]
  \caption{Constraints on $B_{\rm \mu G}$ and $n_{\rm 0}$ based on
    standard DSA predictions (assuming Bohm diffusion) and the ULs
    on the maximum particle energy derived from the \fermi and H.E.S.S. measurements.}
  \label{tab:constraints}  
  \centering
  \renewcommand{\arraystretch}{1.3}
  \begin{tabular}{|c| rcl |}
    \hline
    Scenario  & \multicolumn{3}{c|}{Constraints} \\
     \hline
     Radiative losses $\tau_{\rm rad}$:         &   \multicolumn{3}{c|}{($\tau_{\rm acc}> \tau_{\rm rad}$)}                                  \\
     $\tau_{\rm p-p}$                                   & $B_{\rm \mu G}$&< & ${1.1\times10^{-4}}\,n_0\, E_{\rm max}$                                       \\ [0.5mm]
     $\tau_{\rm br}$                                    & $B_{\rm \mu G}$&< &${1.8\times10^{-4}}\,n_0\, E_{\rm max}$                                        \\ [0.5mm]
     $\tau_{\rm syn}$                                           & $B_{\rm \mu G}$&> &$3400\, E_{\rm max}^{-2}$                \\ [1mm]
    \hline
     Age-limited:        $\tau_{\rm acc}(E_{\rm max}) > {\rm age}$      & $B_{\rm \mu G}$ & $<$ & $1.4\, E_{\rm max}$                                              \\ [1mm]
      \hline
     Size-limited:       $\frac{D(E_{\rm max})}{{\rm v}_{\rm \mathrm{sh}}} > \chi R_{\rm \mathrm{sh}}$      & $B_{\rm \mu G}$ &$<$ & $0.1\,\chi_{0.1}^{-1}\, E_{\rm max}$ \\ [1mm]
    \hline
  \end{tabular} 
  \tablefoot{ $B_{\rm \mu G}$, $n_0$, and the maximum particle energy $E_{\rm
      max}$ are in units of $\mu$G, cm$^{-3}$ and TeV, respectively.
    The shock velocity v$_{\rm sh}$=700~km\,s$^{-1}$ is used as it leads to conservative constraints on $B_{\rm \mu G}$.
    The ratio between the diffusion length of particles at $E = E_{\rm max}$ and
    the shock radius $\chi$ defines the upstream diffusion
    region size, where $\chi_{0.1}=\frac{\chi}{0.1}$ \citep[e.g.][]{zp_2008}. 
    $\tau_{\rm p-p}$, $\tau_{\rm br}$, and $\tau_{\rm syn}$ are the radiative loss timescales for p-p
    collision, Br, and syn processes, $\tau_{\rm acc}$
    and $\tau_{\rm rad}$ are the acceleration and radiative
    timescales, respectively, and $D(E_{\rm max})$ the diffusion
    coefficient at the maximum energy.}
\end{table}

In the age- and size-limited scenarios, the derived ULs on the
magnetic field are lower than both the estimates of
\citet{2012ApJ...759...89H}, based on a simple one-zone modelling of
\PA broadband emission (between 8 and 35~$\mu G$), and those of
\citet{2013A&A...555A...9D} and \citet{arbutina_2012}, based on
  equipartition arguments ($\sim$26--100 $\mu$G). In other words,
  higher magnetic field values would have led to higher E$_{\rm max}$
  and hence to VHE $\gamma$-ray emission from \PA detectable with the
  \hess array. The ULs in Table \ref{tab:constraints} depend on the
  diffusion coefficient, which could depart from the traditional Bohm
  assumption \citep[see \eg][]{2006A&A...453..387P}. However, strong
  deviations from the Bohm regime would be required to make the
  non-detection with \hess compatible with the \fermi detection. In
  particular, for the hadronic scenario in the size-limited case,
  diffusion about two orders of magnitude slower than the Bohm one
  would be needed.  In the loss-limited cases due to p-p and Br
  interactions, lower limits on the density are much higher than
  estimated from IR observations towards the NE rim
  \citep[$\sim$4~cm$^{-3}$,][]{2010ApJ...725..585A}, for any
  acceptable value of the magnetic field. Nonetheless, this density
  estimate, together with a realistic amount of energy in accelerated
  particles of $\sim$(1-4) $\times$10$^{49}$~erg, can account for the
  GeV emission \citep{2012ApJ...759...89H}. The synchrotron limited
  case leads to a lower limit on the magnetic field that is far too high for a
  SNR of age $\sim$4500 yr.  \cite{2012ApJ...759...89H} have treated
  E$_{\rm max}$ as a free parameter in their broadband modelling,
  fixing it to 0.5 (resp. 0.8) TeV in their leptonic (resp. hadronic)
  modelling in order not to violate the \fermi~measurements, but
  without imposing any physical constraint. By applying the same
  reasoning as above with only these \fermi~lower limits on $E_{\rm
    max}$, non-constraining limits on $B_{\rm \mu G}$ and $n_0$ are
  obtained.

  The hypothesis of a break in the particle spectrum of \PA~owing to
  synchrotron losses results in a realistic lower limit on the
  magnetic field ($B_{\rm \mu G} > 70\, E_{\rm break}^{-1/2}$, $E_{\rm
    break}$ in TeV).  However, leptonic-dominated scenarios would
  require an unusually high electron-to-proton ratio (greater than
  0.1), in excess of the observed CR abundances
  \citep{2012ApJ...759...89H}. Radiative breaks due to p-p and Br
  mechanisms imply a constraint on n$_0 \gtrsim$10$^4$~cm$^{-3}$ that is much
  stronger than the density estimates reported in \PA.  Therefore, it
  turns out that none of the known limitations in the simple context
  of a single population of particles continuously accelerated at the
  SNR shock can explain the lack of VHE emission from the \PA SNR,
  except if the diffusion has been proceeding far from the Bohm limit.

  However, if the SNR shock has encountered a MC some time ago, the
  acceleration of particles could have ceased because of ion-neutral
  damping. In such a case, a radiative cutoff would appear at an
  energy for which $\tau_{\rm rad} = \Delta t$, with $\Delta t$ the
  time elapsed since the beginning of the interaction. This would
  imply $n_0 \gtrsim 10^3$~cm$^{-3}$ and $B \gtrsim 50 \,\mu$G for the
  Br/p-p and syn radiative losses, respectively. These values seem to
  be very reasonable for a MC \citep{1999ApJ...520..706C}.

  Alternatively, other scenarios that could explain a break in the HE
  regime deal with particle escape and diffusion in SNRs
  \cite[\eg][]{2010A&A...513A..17O,2012PhPl...19h2901M}. Although
  these spectral breaks are generally observed at energies of
  $\sim$1$-$20~GeV \citep{2009ApJ...706L...1A,2010ApJ...718..348A,
    2010Sci...327.1103A, 2010ApJ...712..459A, 2010ApJ...722.1303A},
  so lower than the constraints presented here, it is not clear whether
  the detection of such breaks in this energy range is entirely due to
  a \fermi~statistical selection effect or not. Some localised regions
  along the \PA outer rim are known to have interacted with dense
  surrounding material \citep[such as the
  BEK,][]{2005ApJ...635..355H}, but due to their very small sizes and
  positions along the SNR rim they may not be representative of the
  bulk of the GeV emission observed with \fermi. The GeV emission is
  more compatible with the (hard) X-ray morphology shown in
  \citet{2013A&A...555A...9D} pointing towards the NE region adjacent
  to a cloud traced in the far-IR domain that either still needs to be hit
  or that has already being shocked by the \PA SNR. Such a cloud interaction
  could be responsible for a break in the HE/VHE \g-ray spectrum
  through the above-mentioned mechanisms, but at somewhat higher
  energies than observed in the more evolved interacting SNRs.

\section{Conclusion}

The H.E.S.S. observations of \PA in the VHE domain reveal an
unexpected lack of emission from the SNR.  The extrapolation from the
\fermi HE power-law spectrum to the VHE domain contrasts with the
absence of VHE emission. The comparison of these two measurements
indicates that a spectral feature (a break or a cutoff) must exist at
energies around a few hundred GeV. By assuming a PL with an
exponential (resp. sub-exponential) cutoff, such a feature should
occur below 450 GeV (resp. 280 GeV) at the 99\% CL. The latter value
provides a conservative UL on any break energy as long as $\Delta
\Gamma \lesssim 1$. In the context of a single population of particles continuously
accelerated at the SNR forward shock through an on-going DSA process, and under the assumption of Bohm diffusion, it is
difficult to reconcile the constraints on the magnetic field and ISM
density derived from the broadband emission modelling
\citep{2012ApJ...759...89H} with those obtained here based on the
predicted maximum/break particle energies. However, multi-wavelength
data suggest that \PA has already interacted with MCs in some
localised regions along the shell and that the NE region coincident
with the bulk of GeV emission is possibly interacting with a far-IR MC
\citep{2013A&A...555A...9D}.  If this is true, the acceleration of
particles could have ceased some time ago, and either a radiative
cutoff or a break of a non-radiative origin could be expected. In the
latter case, the break is expected at somewhat higher energies than
those measured in several SNRs known to be interacting with MCs, which
lie in the 1-20~GeV energy range \citep{2009ApJ...706L...1A,
  2010ApJ...722.1303A, 2010ApJ...718..348A,
  2013Sci...339..807A}. Upcoming observations with the five-telescope
\hess~II will allow the unexplored $\sim$100$-$300 GeV domain, where
this spectral feature is predicted to exist, to be probed for the
first time.

\begin{acknowledgements}
  The support of the Namibian authorities and of the University of
  Namibia in facilitating the construction and operation of
  H.E.S.S. is gratefully acknowledged, as is the support by the German
  Ministry for Education and Research (BMBF), the Max Planck Society,
  the German Research Foundation (DFG), the French Ministry for
  Research, the CNRS-IN2P3, and the Astroparticle Interdisciplinary
  Programme of the CNRS, the U.K. Science and Technology Facilities
  Council (STFC), the IPNP of the Charles University, the Czech
  Science Foundation, the Polish Ministry of Science and Higher
  Education, the South African Department of Science and Technology
  and National Research Foundation, and by the University of
  Namibia. We appreciate the excellent work of the technical support
  staff in Berlin, Durham, Hamburg, Heidelberg, Palaiseau, Paris,
  Saclay, and in Namibia in the construction and operation of the
  equipment. We thank Gloria Dubner for providing the 1.4 GHz
  radio image of Puppis A.
\end{acknowledgements}

\bibliographystyle{aa}  

\bibliography{PuppisA.bib}

\end{document}